\documentclass{aa}  
\usepackage{amssymb,epsfig,graphicx}
\usepackage{rotating}

\usepackage{txfonts}
\begin{document}
   \title{The nature of the dwarf starforming
galaxy associated with GRB\,060218 / SN\,2006aj  \thanks{Based on observations obtained at the ESO VLT under ESO programme 076.A-0737(A).
}}

\author{ K.~Wiersema \inst{1} \and     
S.~Savaglio \inst{2} \and
P.~M.~Vreeswijk \inst{3,4} \and
S.~L.~Ellison \inst{5} \and
C.~Ledoux \inst{3} \and 
S.-C.~Yoon \inst{1} \and 
P.~M\o ller \inst{6} \and
J.~Sollerman \inst{7} \and
J.~P.~U.~Fynbo \inst{7} \and
E.~Pian \inst{8} \and
R.~L.~C.~Starling \inst{1} \and
R.~A.~M.~J.~Wijers \inst{1}
}

\institute{Astronomical Institute `Anton Pannekoek', University of Amsterdam,
	  Kruislaan 403, 1098 SJ Amsterdam, the Netherlands
	  \and
	  Department of Physics and Astronomy, Johns Hopkins University, 
	  Baltimore, MD 21218, USA
	  \and
	  European Southern Observatory, Alonso de C\'{o}rdova 3107, 
	  Casilla 19001, Santiago 19, Chile
	  \and
	  Departamento de Astronom\'ia, Universidad de Chile, Casilla 36-D, Santiago, Chile
	  \and
	  Department of Physics and Astronomy, University of Victoria, Elliott Building, 3800 Finnerty Rd, Victoria, BC, V8P 1A1, Canada
	  \and
	  European Southern Observatory, Karl-Schwarzschild-strasse 2, D-85748 Garching bei M\"{u}nchen, Germany
	  \and
	  Dark Cosmology Centre, Niels Bohr Institute, University of
	  Copenhagen, \mbox{Juliane Maries Vej 30, 2100 Copenhagen, Denmark}
	  \and
	  INAF, Astronomical Observatory of Trieste, I-34131 Trieste, Italy
}
   \offprints{kwrsema@science.uva.nl}

   \date{Received 2006; accepted 2006}

 
  \abstract
   {We present high resolution VLT UVES and low resolution FORS optical spectroscopy of
supernova 2006aj and its host galaxy, associated with the nearby ($z = 0.03342$) gamma-ray burst
GRB\,060218. This host galaxy is a unique case, as it is one of the few nearby GRB host galaxies known, and 
it is only the second time high resolution spectra have been taken of a nearby GRB host galaxy (after GRB\,980425).
}
   {The resolution, wavelength range and S/N of the UVES spectrum combined with low resolution FORS spectra allow a 
   detailed analysis of the circumburst and host galaxy environments. } 
   {We analyse the emission and absorption lines in the spectrum, combining the high resolution UVES spectrum 
   with low resolution FORS spectra and find
   the metallicity and chemical abundances in the host.
   We probe the geometry of the host by studying the emission line profiles.}
   {Our spectral analysis shows that the star forming region in the host is metal poor with 12 + log(O/H) = $7.54^{+0.17}_{-0.10}$ ($\sim0.07\,Z_{\sun}$), 
   placing it among the most metal deficient subset of emission-line galaxies. It is also the lowest
   metallicity found so far for a GRB host from an emission-line analysis.
    Given the stellar mass of the galaxy of $\sim10^7$M$_{\sun}$ and the SFR$_{\rm H \alpha}$ = 0.065 $\pm$ 0.005 M$_{\sun}$\,yr$^{-1}$, 
   the high specific star formation rate indicates an age for the galaxy of less than
   $\sim200$ Myr.
The brightest emission lines are clearly asymmetric and are well fit by two Gaussian 
   components separated by $\sim 22$ km s$^{-1}$. We detect two discrete \ion{Na}{I} and \ion{Ca}{II} absorption components at the same redshifts 
   as the emission components.
   We tentatively interpret the two components as arising from
two different starforming regions in the host, but high resolution
imaging is necessary to confirm this. 
  }
   {}

\titlerunning{The nature of the host galaxy of GRB\,060218 / SN\,2006aj}
   \authorrunning{Wiersema et al.}
   \maketitle
%

\section{Introduction}

Long-duration gamma-ray bursts (GRBs) are widely accepted to be related to core-collapse supernovae:  
clear supernova signatures are seen in the afterglow spectra of low redshift GRBs
(e.g. Stanek et al.~2003; Hjorth et al. 2003; Pian et al.~2006).
Dedicated surveys of GRB hosts suggest that GRBs occur
preferentially in low mass, subluminous, blue star-forming
galaxies (e.g.~Chary et al.~2002; ~Fruchter et al.~1999; Le Floc'h et al.~2003).
The GRBs are often located within UV-bright parts of their
hosts, where star formation takes place (Bloom et al.~2002; Fruchter et
al.~2006), and are shown to be more concentrated towards the brightest regions of their hosts than are, in general, 
core-collapse supernovae (Fruchter et al.~2006).
 
GRB host galaxies are not selected through their
luminosities or colours, but merely by the fact that a GRB has been
detected. This could potentially provide an unbiased sample of starforming galaxies, which may be used to
study star formation in the Universe (see e.g.\ Jakobsson et al.~2005). 
The significant increase in detection rate and localisation of 
long GRBs through the successful operation of {\em Swift} in principle 
permits the study of a large and uniformly selected sample of GRB host galaxies.
However, to date the sample of {\em spectroscopically} studied GRB host galaxies is small ($\sim$30 cases) and 
may suffer from several selection biases, due to their faintness.

One of the key properties needed to understand GRB progenitors and their environments is the metallicity distribution (e.g.~Langer \& Norman 2006; 
Yoon, Langer \& Norman 2006): the popular collapsar model for long GRBs requires low metallicity progenitors.
Metallicities for GRB hosts can be determined through afterglow spectroscopy and through host galaxy spectroscopy. 
Absorption lines of \ion{H}{I} and heavy metals have provided metallicities along GRB sight-lines in the redshift
interval $2<z<6.2$. 

Host galaxy spectroscopy can provide metallicities for the galaxy as a whole, but relies on the detection of the nebular emission lines, which 
is difficult at $z \gtrsim 1$. 
There are only a few GRB host galaxies that are
bright enough to permit a direct abundance analysis through an
electron temperature ($T_e$) determination (GRBs\,980425, 020903, 031203; Prochaska et al.\ 2004, Hammer et al.\ 2006).
It is only possible to study the metallicities of higher redshift and fainter hosts through secondary
metallicity indicators using bright (nebular) lines (e.g.\ the R23 method), using empirical correlations between the fluxes of certain emission lines and metallicity.

On the 18th of February 2006 a bright, nearby GRB
was discovered by {\it Swift}.  The proximity ($z \sim 0.0334$, Mirabal
\& Halpern 2006, the second closest GRB) of this GRB triggered a large follow-up campaign by
several groups,
which provided a unique opportunity to unveil the nature of a
faint nearby galaxy associated with a GRB event.
GRB\,060218 was found to be unusually long in duration ($T_{90} = 2100 \pm 100$ s, Campana et al.~2006) and of relatively
low luminosity ($E_{\gamma, {\rm iso}} = 6.2 \pm 0.3 \times 10^{49}$ erg, Soderberg et al.~2006). Its prompt emission was soft,
placing it in the class of X-ray flashes (XRFs).
A bright supernova (designated SN\,2006aj) was clearly associated with this event
which was studied with 
very high spectral and time resolution over a wide range of wavelengths, from X-rays to
radio (e.g. Campana et al.~2006; Pian et al.~2006; Mazzali et al.~2006; Soderberg et al.~2006). 
The host was found to be a small galaxy with an irregular morphology.

In this paper we study the host of GRB\,060218 through a high resolution VLT Ultraviolet and Visual 
Echelle Spectrograph (UVES; Dekker et al.~2000) spectrum, taken
around the peak magnitude of the supernova.
This spectrum shows a variety of well resolved emission
lines, associated with ionized gas in the starforming
regions in the host. The neutral gas of the ISM is probed by the
detection of a few narrow absorption lines. Low resolution FORS spectra (described in Pian et al.~2006) are used to 
study the fluxes of emission lines that fall outside the spectral coverage of the UVES spectrum.

Several papers have already been published on GRB\,060218, its associated supernova, and the host.  
However, on the metallicity of the host there
is a wide range of reported values. Modjaz et al.~(2006) report a metallicity of 0.15 Z$_{\sun}$, 
Mirabal et al.~(2006) derive 0.46 Z$_{\sun}$,
while Sollerman et al.~(2006) mention that the abundance is below Solar, but that 
the exact value is unconstrained from the strong emission lines. These values are derived from different spectra, but 
the spread is largely due to internal scatter in empirical, secondary calibrators that are used, 
and the limited range of metallicities and diagnostic emission line ratios for which these secondary calibrators 
are valid (see e.g. Ellison \& Kewley 2006). 
Since the derived metallicities are frequently used 
to draw conclusions on important issues such as progenitors (e.g.~Sollerman et al.~2005) or the rate of 
nearby GRBs (Stanek et al.~2006; Wolf \& Podsiadlowski 2006), it is clearly of great interest to pin down these uncertainties.

The paper is organized as follows: In \S2 we describe the observations. 
In \S3 we calculate the metallicity of the dominant starforming region(s) in the 
host galaxy and relative abundances of Ne and N, as well as the star formation  
rate from optical and radio fluxes. In \S4 we analyse the discrete velocity components
in the host through emission and absorption lines. We discuss the use of metallicity values found from 
secondary metallicity calibrators in \S5. In \S6 we discuss the implications that a future detection of Wolf-Rayet star 
signatures in the host of GRB\,060218 may have on single star GRB progenitor models. 
 
Throughout this paper we use
the cosmological parameters H$_{0}$ = 70 km s$^{-1}$/Mpc, $\Omega_{\rm M}$ =
0.3 and $\Omega_\Lambda$ = 0.7.

\section{Observations \label{obssection}}

GRB\,060218/SN\,2006aj was observed with ESO VLT {\it Kueyen} (UT 2) on March 4, 2006,
roughly at the time of maximum light of the supernova (around March 1, e.g.~Sollerman et al.~2006).
The UVES observation
started at 00:30 UT, for a total exposure time of 2100 seconds. 
The magnitude of SN\,2006aj was $V \sim 17.6$ at the time of observation.
The airmass was high, averaging 2.5, and the seeing measured from the 2D spectrum is 1.1 arcsec.
The UVES-setup spectral resolution is $R \sim 46000$ (FWHM $\simeq6.5$ km s$^{-1}$).  
The slit width was set at 1 arcsec, corresponding to about 7 kpc physical size at
the distance of the GRB. The exposure was performed at the parallactic angle.
The spectrum (wavelength ranges 3285 - 4527 \AA, 4621 - 5598 \AA~and
5676 - 6651\AA) was reduced in the standard fashion using MIDAS and IRAF routines.

The {\em L.A.Cosmic} program
(specifically the {\em lacos\_spec} routine, Van Dokkum 2001) was
used to remove both point- and irregular shaped cosmic ray hits from
the 2D spectrum before extraction. 
The spectrum
was dispersion corrected, and flux calibrated by archived response
functions. An air to vacuum conversion
and a heliocentric correction (+28 km s$^{-1}$) were applied to the spectrum.
The resulting spectrum was compared with a (quasi)-simultaneous FORS
spectrum taken 20 minutes after the UVES spectrum, which is flux calibrated through a standard star observation
and through simultaneous photometry in $B, V, R$ at VLT FORS2 (Pian et
al.~2006), using the magnitude to flux conversions from Fukugita,
Shimasaku \& Ichikawa (1995).
To the UVES spectrum a multiplication factor of 1.625 was applied to match the well calibrated FORS flux
values (compensating for the slit loss between the FORS and UVES spectra). 

\begin{figure}
   \centering
   \includegraphics[width=8.5cm]{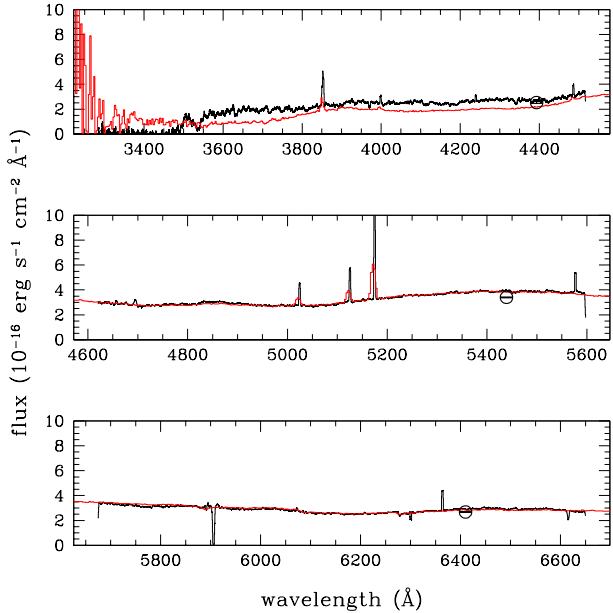}
      \caption{The UVES spectrum shown in black, with overlaid in red the low resolution FORS spectrum. For presentation purposes the
      UVES spectrum has been smoothed to the pixelscale of the FORS spectrum. 
      The points denote $B,V,R$ VLT FORS photometry from the same night (Pian et al.~2006). The widths of the broadband filters are not shown.	
	     }
	\label{UVEScompplot}
   \end{figure}

We find a good match between the UVES and FORS spectra and photometry
in the red end (above $\lambda \sim$4500 \AA), as shown in Figure \ref{UVEScompplot}. 
In the blue end ($\lambda \lesssim 4500$ \AA) the UVES continuum flux is 
slightly higher than the FORS continuum flux. 
The prime reason is likely
the very high airmass at which both spectra have been taken, making the flux calibration of both
the UVES and FORS spectrum at the blue end more uncertain.
We decide to 
not alter the flux calibration, but warn that the fluxes of emission lines below $\sim$4500 \AA~have a small 
additional uncertainty. 
This uncertainty does not significantly affect the metallicity results of this paper,
because the electron temperature uncertainty is dominated by the uncertainty in the 
flux of [\ion{O}{III}] $\lambda$4364 from the FORS spectra. The host galaxy emission line flux ratios found in the
UVES spectrum agree, within the errors, with those found from the combination of the FORS spectra.

A Galactic extinction correction was performed using $E(B-V)_{\rm MW}$
= 0.142 (Schlegel, Finkbeiner \& Davis 1998), 
assuming a Galactic extinction law
$A_{\lambda}/A_{V}$ expressed as $R_{V} = A_{V}/E(B-V)$ (Cardelli et
al.~1989), and $R_{V} = 3.1$. This value is slightly higher than the
extinction derived by Guenther et al.~(2006), who 
used the Galactic sodium lines to find $E(B-V)_{\rm MW} = 0.127$. 
Given that the systematic error in the conversion of the
equivalent widths of \ion{Na}{} to $E(B-V)$ is poorly known, we choose to use $E(B-V)_{\rm MW}$ = 0.142 mag. 

\begin{figure}
   \centering
   \includegraphics[width=8.5cm]{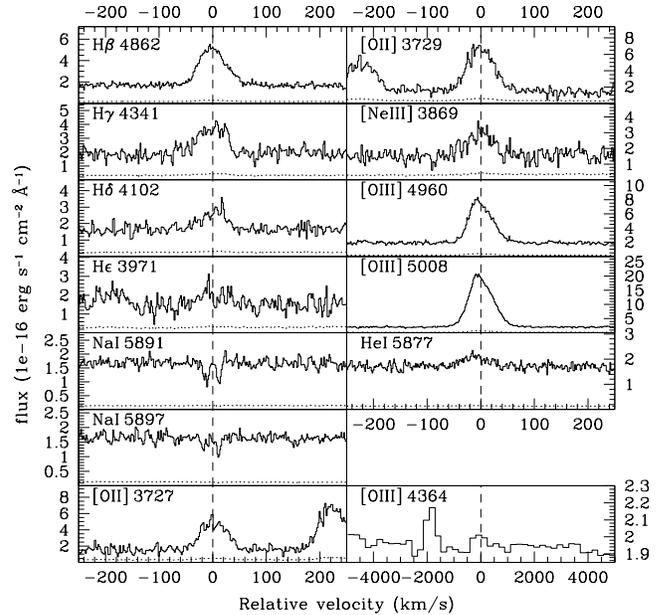}
      \caption{Emission and absorption lines detected in the UVES spectrum that were used to derive the host properties. 
               The error spectra are plotted with dotted lines. The vertical dashed lines denote the mean redshift of the host. The bottom
	       right panel shows the [\ion{O}{III}] $\lambda$4364 line from the FORS spectrum. 
              }
         \label{UVESspecplot}
   \end{figure}

A selection of lines from the spectrum and the
associated error spectra are shown in Figures~\ref{UVESspecplot} and \ref{fig1_sa}.
Although the spectrum is dominated by the SN spectrum ($V\approx17.6$ vs the host $V\approx20.2$),
several bright emission lines from the host galaxy are detected and resolved in the UVES
spectrum (Fig.~\ref{UVESspecplot}, Table~\ref{linestable}). 
On the other hand, the detected absorption lines, from both the host galaxy and Milky Way,
are narrower than the UVES resolving power.  The
highest signal-to-noise emission lines provide a heliocentric mean redshift $z = 0.03342(2)$ (Pian et al.~2006), where we
adopt a conservative error on the redshift due to poorly known systematic effects of the spectrograph (e.g. the 
centering of the object in the slit).

The analysis of the emission lines was done using the IRAF software
packages, mainly using the {\em splot} routines. As the emission lines are clearly
asymmetric (see \S4.1), fluxes are measured using the numerical integration
method ({\em e} in {\em splot}), and not the standard Gaussian
fitting. The errors in the line fluxes are generally dominated by the
uncertainty in the continuum level and are given at the 1$\sigma$ level.

\section{General host galaxy properties}
\subsection{Metallicity}
\subsubsection{Reddening}
Before we can use the emission lines in the FORS and UVES spectra to derive the host properties, we need to correct for the dust
extinction intrinsic to the host. 
We use the Balmer line fluxes to measure a 
value for the Balmer decrement, assuming case B recombination (e.g.~Osterbrock 1989; Izotov et al.~1994). 
The detection of several of the Balmer lines in the UVES spectrum, see Table \ref{linestable}, 
provides a good constraint on the extinction value.
We assume intrinsic Balmer line ratios at 20000 K, and compute flux ratios of the detected Balmer lines. 
We caution that H$\delta$ and H$\epsilon$ are located in the blue end of the UVES spectrum, where we see a 
slight discrepancy in flux calibration between
the FORS spectrum of March 4 and the UVES spectrum (the UVES spectrum having a slightly higher mean continuum flux). 
We find that the Balmer line ratios from the UVES spectrum are all (H$\beta$/H$\gamma$ to H$\beta$/H$\epsilon$) 
consistent with the theoretical case B recombination values. The FORS spectra are taken at low resolution ($R \sim300$), making the 
extinction derivation based on the H$\alpha$/H$\beta$ ratio from these
data less reliable.
We test various intrinsic stellar Balmer absorption strengths, but find no evidence for internal reddening from the Balmer decrement in the 
UVES spectrum, with upper limit $E(B-V) \lesssim 0.03$. However this value may be influenced by the uncertain
flux calibration of the blue end of the spectrum, see section \ref{obssection}. The H$\beta$ and H$\gamma$ lines are located 
in an area of the spectrum where the FORS and UVES continuum fluxes are more in agreement. 
The ratio of the H$\beta$ and H$\gamma$ lines is consistent with case B recombination values, $E(B-V) \lesssim 0.04$ (1 sigma).
In Guenther et al.~(2006) a reddening of $E(B-V)$ = 0.042 $\pm$ 0.003 is derived from the strengths of the sodium absorption lines in the host, 
and used by Pian et al.~(2006) to account for reddening in the host. 
We choose to use the more common method of accounting for extinction by using the Balmer decrement.
The resulting net extinction (Galactic + host) is close to the value applied by Pian et al.~(see also Sollerman et al. 2006 for a discussion) and
does not affect further analysis, as the uncertainties in the line ratios are dominated by the uncertainties in their fluxes.
We further note that the high electron temperature seen in this source (see section \ref{tesection}) can make collisional excitation of neutral 
hydrogen important, which can mimic reddening and affect the Balmer decrement derived reddening. We can not reliably evaluate this
effect, as we do not measure a significant Balmer decrement.

We use various nebular line flux ratios 
to evaluate the possibility of an Active Galactic Nucleus (AGN) contribution to the excitation 
of the nebular lines.
Kauffmann et al.~(2003) refine the popular line ratio diagnostic [\ion{O}{III}] $\lambda$5008\,/\,H$\beta$ vs [\ion{N}{II}]\,/\,H$\alpha$, by analyzing a large sample of galaxies
from the SDSS. They empirically define the demarcation between starburst galaxies and AGN as follows: a galaxy is AGN dominated if
\begin{equation}
\log([\ion{O}{III}]\lambda 5008/{\rm H}\beta) > 0.61/(\log([\ion{N}{II}]/{\rm H}\alpha) - 0.05) + 1.3.
\end{equation} 
Kauffmann et al.~(2003) use extinction corrected fluxes, but note that these line ratios are relatively insensitive to extinction effects.
We use the flux values measured from the FORS spectra, 
 [\ion{O}{III}] $\lambda$5008\,/\,H$\beta$ = 4.55 $\pm$ 0.28
and [\ion{N}{II}]\,/\,H$\alpha$ = 0.06 $\pm$ 0.01.  These flux ratios place the host galaxy comfortably in the locus of
actively starforming galaxies.
Hence we assume that the host galaxy nebular emission line excitation 
is not dominated by (non-thermal) AGN emission.
\subsubsection{Electron density and temperature \label{tesection}}
We make the standard two zone model assumption for the \ion{H}{II} region(s) from which the emission lines originate, 
consisting of a low and intermediate temperature region (e.g.~Osterbrock 1974). 
The [\ion{O}{II}]~$\lambda\lambda$3726 and 3729 \AA~lines (Figure \ref{UVESspecplot}) are closely linked to collisional excitation and de-excitation
at the typical temperatures of star forming regions ($\sim$ 10$^4$ K). The electron density ($n_e$) in the low temperature region 
can be derived from the ratio of the 
fluxes of these two lines (Osterbrock 1974). The ratio of the line fluxes $\lambda$3726\,/\,$\lambda$3729 
approaches $\frac{4}{6}$ in the $n_e \rightarrow 0$ limit. In the limit $n_e \rightarrow \infty$ the flux ratio approaches 
2.86 (collisional excitation and de-excitation dominate over radiative transitions, forming a Boltzmann population ratio). 
As shown in Fig.~2, in the UVES spectrum the [\ion{O}{II}] doublet is clearly resolved, 
and its flux ratio is $0.62 \pm 0.05$. 
This number is consistent with the low density limit, implying that collisional de-excitation is not important 
for the fluxes of the forbidden lines. 
The resolved [\ion{O}{II}] doublet has also been observed in two other GRB host galaxy spectra, 
GRBs\,990506 and 000418, which have values of [\ion{O}{II}] $\lambda$3726\,/\,$\lambda$3729
of 0.57 $\pm$ 0.14 and 0.75 $\pm$ 0.11, respectively (Bloom et al.~2002). These two detections imply low values for 
$n_e$ similar to GRB\,060218.

A different doublet used frequently as a density diagnostic is the [\ion{S}{II}] $\lambda\lambda$6717, 6731 doublet.
The line ratio [\ion{S}{II}] $\lambda$6717\,/\,$\lambda$6731 approaches 1.5 when $n_e \rightarrow 0$, and
0.44 above $n_e \sim10^4$ cm$^{-3}$. These lines are redshifted out of the UVES wavelength range, but 
are detected in the FORS spectra. We find [\ion{S}{II}] $\lambda$6717\,/\,$\lambda$6731 = 1.0 $\pm$ 0.6, which is therefore not useful to
discriminate between the high and low density regimes.
Prochaska et al.~(2004) find a ratio of $1.19 \pm 0.09$ for the host of GRB\,031203, corresponding to $n_e \sim 300$ cm$^{-3}$.
We assume the relatively low values of $n_e = 100$ cm$^{-3}$ for our analysis, following e.g.~Skillman et al.~(1994); Izotov et al.~(2006b).

We estimate the electron temperature ($T_e $) in the intermediate temperature region from the ratio of 
[\ion{O}{III}] nebular and auroral fluxes. 
The auroral [\ion{O}{III}]~$\lambda$4364 is not significantly detected in the UVES spectrum - it is located in a noisy region close to the 
gap in between the wavelength ranges, see \S2. 
 
From the FORS spectrum we find the flux ratio 
[\ion{O}{III}] ($\lambda$4959 + $\lambda$5008)\,/\,$\lambda$4364 = 29.1 $\pm$ 6.3. 
The relatively large error is mainly due to the rather large uncertainty in the 
[\ion{O}{III}] $\lambda$4364 flux value. 
We use the electron density assumed above, which we take to be the same in both the
low- and intermediate temperature regions (see e.g.~Osterbrock 1974), and find an electron temperature $T_e ({\rm O}^{2+})$
in the intermediate temperature region of
2.48\,$^{+0.5}_{-0.3}$ $\times 10^4$  K. 
This is high when compared to that of the host of GRB\,031203 ($T_e ({\rm O}^{2+}) \sim 13400$ K, Prochaska et al.~2004), as
shown in Table 1.
Comparably high values of $T_e ({\rm O}^{2+})$ have been observed in the recent discovery of two extremely
low metallicity galaxies by Izotov et al.~(2006b). 
This very high temperature and low density suggests a low oxygen abundance for the host of GRB\,060218, since the main nebular cooling 
is done through oxygen forbidden line emission. 

Especially at low metallicity there can be large differences in electron temperature between the low and high temperature zones.
Due to a lack of detected lines that can be used as temperature indicators in the low temperature region 
(the [\ion{O}{II}] $\lambda\lambda$7320, 7331 are redshifted out of
the UVES coverage and are too faint for the FORS spectrum), we follow the recipe by Izotov et al.~(2006a), 
\begin{equation}
T_e \left({\rm O}^{+}\right) = -0.577 + T_e \left({\rm O}^{2+}\right) \times \left(2.065 -0.498 T_e \left({\rm O}^{2+}\right)\right),
\end{equation}
where $T_e$ is in units of 10$^4$K. We find $T_e ({\rm O}^{+}) = 1.5 ^{+0.1}_{-0.2} \times 10^4$ K, using the
approximation for a low metallicity environment (12 + log O/H $\sim$ 7.2, see Izotov et al.~2006a). 
When we assume an intermediate metallicity (12 + log O/H $\sim$ 7.6), we find
$T_e ({\rm O}^{+}) = 1.3  ^{+0.2}_{-0.5} \times 10^4$ K.
We adopt the low metallicity value for now, and note that the prescription by Pagel et al.~(1992) gives a similar value of
$T_e ({\rm O}^{+}) = 1.66 ^{+0.1}_{-0.08} \times 10^4$ K. 
The conversion above from $T_e \left({\rm O}^{2+}\right)$ to $T_e \left({\rm O}^{+}\right)$ is derived through sequences of photoionization
models (Stasi\'{n}ska \& Izotov 2003). Izotov et al.~(2006a) show by comparing this conversion with direct measurements
of $T_e \left({\rm O}^{2+}\right)$ and $T_e \left({\rm O}^{+}\right)$ that the models agree with the measurements. 
There is a significant scatter of direct measurements of 
$T_e ({\rm O}^{+})$ and $T_e ({\rm O}^{2+})$ with respect to the model predictions, which is probably mainly due to the large uncertainties
in the measurements of $T_e ({\rm O}^{+})$ (Izotov et al.~2006a). We do not include an additional uncertainty
in the following analysis to account for this, but note that the small errors reported here should not be overinterpreted.
We assume the $T_e ({\rm O}^{+})$ to also be valid for \ion{N}{II} and \ion{S}{II}, which have comparable ionisation potentials. 
\subsubsection{Oxygen abundance} 
We use the equations from Izotov et al.~(2006a) to derive ionic abundances from the derived electron densities and temperatures.
To find the oxygen abundance, O/H,  we sum the O$^{2+}$\,/\,H$^{+}$ and O$^{+}$\,/\,H$^{+}$ ion abundances, assuming the O$^{3+}$ abundance is
negligible since no high temperature lines (e.g. \ion{He}{II}) were detected. We take the line fluxes from the UVES spectrum, and
find O$^{2+}$\,/\,H$^{+} = 1.72 \pm 0.45 \times 10^{-5}$ and O$^{+}$\,/\,H$^{+} = 1.78 ^{+1.2}_{-0.32} \times 10^{-5}$.
The errors include the uncertainties on the line flux ratios and electron temperature. 
The ratio O$^+$\,/\,(O$^+$ + O$^{2+}$) $>$ 0.1, confirming that the excitation is too low for O$^{3+}$ to be important, as shown by 
photoionization models by Izotov et al.~(2006).
The large uncertainty in the O$^{+}$\,/\,H$^{+}$ abundance is caused by the relatively large error in the $T_e ({\rm O}^{+})$.
The total oxygen abundance is O\,/\,H = O$^{2+}$\,/\,H$^{+}$ + O$^{+}$\,/\,H$^{+}$ = 3.50 $^{+1.65}_{-0.77} \times 10^{-5}$, or
12 + log(O/H) = 7.54 $^{+0.16}_{-0.1} $,  or $\simeq 0.07 Z_{\sun}$, 
assuming log(O/H)$_{\sun}$ + 12 = 8.69 (Allende Prieto et al.~2001). 
This value places the host of GRB\,060218 among the most metal deficient subset of emission line galaxies in the local universe
(e.g. Izotov et al.~2006a, Lee et al.~2006). 
It is also the lowest metallicity found so far for a GRB host from emission line analysis (absorption line metallicities from several 
afterglows show lower line of sight metallicities, down to $\sim$0.01 $Z_{\sun}$ for GRB\,050730, Starling et al.~2005; Chen et al.~2005).
We note that in an independent analysis Kewley et al.~(2006) found a similar value for the metallicity of the host of GRB\,060218 of
12 + log(O/H) $\sim$ 7.6.

\subsection{Relative element abundances}

Accurate emission-line abundances have been derived for a small sample of GRB host galaxies. A notable example is the spectrum of the host of GRB\,031203, for which a 
solar abundance pattern was established (Prochaska et al.~2004). 
We use the host emission lines measured in the UVES and FORS spectra 
to gain an insight into the abundance pattern in the host of GRB\,060218.

The detection of the forbidden [\ion{Ne}{III}] lines allows us to derive a Ne abundance, using 
the values for $n_e$ and $T_e ({\rm O}^{2+})$ found in \S3.1.
We use the [\ion{Ne}{III}]~$\lambda$3869\,/\,H$\beta$ flux ratio from the UVES spectrum and
find Ne$^{2+}$\,/\,H$^{+} = 2.9 \pm 1.2 \times 10^{-6}$, and Ne$^{2+}$\,/\,O$^{2+}$ = 0.17 $\pm$ 0.11, which is consistent with
other low metallicity H\,II regions, e.g. in I Zw 18 (where Ne$^{2+}$\,/\,O$^{2+} \sim 0.13$, see Table \ref{proptable}).
To derive the Ne/H abundance we need an ionisation correction function (ICF) for which we use the parametrization by Izotov et al.~(2006a).  
We find ICF (Ne$^{2+}$) = 1.11  
and Ne\,/\,H = 3.3 $\pm$ 1.3 $\times 10^{-6}$.
  
The [\ion{S}{III}] $\lambda$6312 line is not detected in the UVES or FORS spectra and
 [\ion{S}{III}] $\lambda\lambda$9532, 9069 are redshifted out of both the UVES and FORS coverage.  
We will therefore only derive a value for the ionic S$^{+}$ abundance, and give an upper limit for the total sulphur abundance.
We use $T_e ({\rm O}^{+})$ and find S$^{+}$\,/\,H$^{+}$ = $ 2.7^{+2.0}_{-1.0} \times 10^{-7}$. 
For the limit on S$^{2+}$\,/\,H$^{+}$ we transform 
$T_e ({\rm O}^{2+})$ to $T_e ({\rm S}^{2+})$ through the recipe of Izotov et al.~(2006a),
and use the upper limit on the [\ion{S}{III}] flux from the UVES
spectrum to find  S$^{2+}$\,/\,H$^{+} < 2.2 \times 10^{-6}$. We calculate an ICF(S$^{+}$+ S$^{2+}$) of $\sim$1 which allows us to 
set the not particularly constraining limit $\log({\rm S}/{\rm O}) < -1.1$ (Solar value is $\log({\rm S}/{\rm O}) = -1.50$, Lodders 2003), which is consistent
with the observed trend for S to follow Solar (S/O) ratios independent of (O/H) for low metallicity \ion{H}{II} regions.

The [\ion{N}{II}] $\lambda$6584 line is redshifted out of the UVES range, but is detected in the FORS spectrum.
The weaker [\ion{N}{II}] $\lambda$6548 line is not significantly detected in the FORS spectrum, with a 3$\sigma$ upper limit on the flux of $\sim 8 \times$
10$^{-17}$ erg s$^{-1}$ cm$^{-2}$. We use the fixed flux ratio $\lambda$6584\,/\,$\lambda$6548 = 2.9 (Osterbrock 1989) and
$T_e ({\rm O}^{+})$, and find N$^{+}$\,/\,H$^{+}$ = $1.4 ^{+0.8}_{-0.4} \times 10^{-6}$.

\begin{figure}
   \centering
   \includegraphics[width=8.5cm]{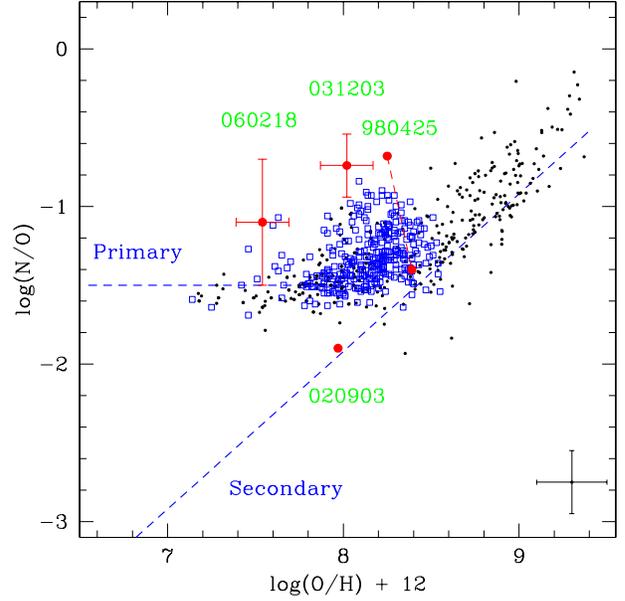}
      \caption{Measurements of log(N/O) and 12 + log(O/H) from a sample of galaxies from the SDSS-DR3 (Izotov et al.~2006) are shown with open squares,
               together with points from Lopez \& Ellison (2003) (small points), for which typical
               uncertainties are indicated in the bottom right corner.  
   	       The blue dashed lines show the tracks where primary and secondary nitrogen dominate the
	       N/O ratio.
	      Published GRB host galaxy measurements are overplotted:    
	      GRB\,060218 (this work), GRB\,031203 (Prochaska et al.~2004), GRB\,980425 (the dashed line connects the
	      value for the SN region with the higher log(N/O) and the general host value, see \S3.2, Hammer et al.~2006) and GRB\,020903 (Hammer et al.~2006).
			       }
	 \label{nitrogen}
   \end{figure}

To calculate N/H from N$^{+}$\,/\,H$^{+}$ we correct for ionisation using ICF (N$^{+}$) = 1.98, and find
N/H = 2.8 $^{+1.6}_{-0.8} \times 10^{-6}$, see Figure~\ref{nitrogen}. The ratio N$^{+}$\,/\,O$^{+}$ = 0.08 $^{+0.07}_{-0.05}$ is comparable to, though slightly above,
the ratio for I Zw 18 of N$^{+}$\,/\,O$^{+}$ $\sim$ 0.03.
We note that here we compare the abundances of two elements using two different spectra (UVES and FORS)
taken under different conditions, and the uncertainty on log(N/O) is likely underestimated. 
Nevertheless the  N/O ratio does not significantly deviate from the observed trend of
low metallicity galaxies to have log(N/O) $\sim -1.5$ (e.g.~Lopez \& Ellison 2003; Izotov et al.~2006).

Hammer et al.~(2006) observed the host of GRB\,980425 / SN\,1998bw with significant spatial resolution 
(owing to the large spatial extent of this host galaxy), and
found a high ratio $\log({\rm N}/{\rm O}) = -0.6$ from a $T_e$ abundance analysis at the region at which the GRB / SN took place, which 
corresponds to almost twice the Solar value. This is unexpected at the measured metallicity, see Figure~\ref{nitrogen}. 
Prochaska et al.~(2004) find a similarly high value of $\log({\rm N}/{\rm O}) = -0.74 \pm 0.2$ dex in 
their spectrum of the host of GRB\,031203, at a metallicity of log(O/H) + 12 = 8.1. GRB\,020903 shows a value 
more in line with expectation from its metallicity (see Fig.3).
In the case of GRB\,060218 the uncertainty on the (N/O) ratio is too high to exclude a deviation from the expected value at the metallicity of the host. 
A possible high N/O value can be explained by a variety of reasons. If $T_e$ has been highly overestimated the metallicity would decrease, 
moving the points in Fig.~\ref{nitrogen} to the left. Physical reasons for an enhanced N/O ratio may be e.g.\ a contribution of shock heating to the line
emission or a chemical evolution effect: Hammer et al.~(2006) explain the higher N/O ratio at the locus of GRB\,980425 by a larger N yield of a GRB
progenitor or SN remnants.

The hosts of GRBs\,980425, 031203, 020903 and 060218 form a sequence in metallicity, from the
metallicities where we expect both primary and secondary nitrogen production to play an active role (the host of GRB\,980425) 
to where primary N is expected to dominate (as in the host of GRB\,060218).
This makes the GRB\,060218 host an interesting candidate for deep spectroscopy to obtain a more accurate N/O when the SN has fully faded.

\begin{table*}[t] 
\begin{center} 
\begin{tabular}{llllll} 
\hline 
Property & GRB\,060218 & GRB\,031203  & I Zw 18 NW & I Zw18 SE  \\
\hline 
$T_{e}(\ion{O}{III}) $ & 2.48 $ ^{+0.5}_{-0.3}  \times 10^4$  K & 13400 $\pm$ 2000&19780 $\pm$ 640&19060 $\pm$ 610\\
$T_{e}(\ion{O}{II}) $ & $1.5 ^{+0.1}_{-0.2}\times 10^4$  K & 12900&15620 $\pm$ 470&15400 $\pm$ 460\\
O$^{2+}$/H$^{+}$  &$1.72 \pm 0.45 \times 10^{-5}$  &-& $1.216 \pm 0.09 \times 10^{-5}$&$1.106 \pm 0.082 \times 10^{-5}$\\
O$^{+}$/H$^{+}$&$1.78 ^{+1.2}_{-0.32} \times 10^{-5}$ &-&$0.179 \pm 0.014 \times 10^{-5}$ &$0.403 \pm 0.031 \times 10^{-5}$\\
O/H &3.50 $^{+1.65}_{-0.77} \times 10^{-5}$ & - &$1.465 \pm 0.092 \times 10^{-5}$ & $1.523 \pm 0.088 \times 10^{-5}$\\
12 + log (O/H) &7.54 $^{+0.16}_{-0.1} $ & 8.02 $\pm$ 0.15&$7.166 \pm 0.027$&$7.183 \pm 0.025$\\
N$^{+}$/H$^{+}$ &$1.4 ^{+0.8}_{-0.4} \times 10^{-6}$ & -& - &$1.074 \pm 0.084 \times 10^{-7}$ \\
ICF (N) &1.98 & - &- & 3.78\\
log (N/O) &$-1.1 \pm 0.4$& -0.74 $\pm$ 0.2 & -&$-1.574 \pm 0.06$\\
Ne$^{2+}$/H$^{+}$ &$2.9 \pm 1.2 \times 10^{-6}$& -& $ 1.91 \pm 0.16 \times 10^{-6}$&$1.83 \pm 0.15 \times 10^{-6}$\\
ICF (Ne) &1.11& -&  1.20&1.28\\
log (Ne/O) &$-0.8 ^{+0.2}_{-0.4} $ & -0.85 $\pm$ 0.2&$-0.803 \pm 0.053$&$-0.781 \pm 0.051$\\
\hline
\end{tabular} 
\end{center} 
\caption{Table of properties of the hosts of GRB\,060218 (this work) and GRB\,031203 (Prochaska et al.~2004) and a comparison to the northwest and 
southeast regions of I Zw 18 (Izotov et al.~1999).  
\label{proptable}} 
\end{table*}

The measured nitrogen, oxygen and neon abundances derived for
the host galaxy of GRB\,060218 are shown in Table~\ref{proptable}. We note that these are not spatially resolved. Izotov et al.~(1999) noted that in the case of I Zw 18 a gradient in electron
temperature can be seen, with the highest temperatures in the regions where WR stars are found. These differences in temperature 
are associated with significant differences in (oxygen) abundance (with factors up to $\sim$1.4, Izotov et al.~1999). 
This gradient may be due to oxygen enrichment by starforming clusters, and incomplete mixing in the galaxy.  
Without spatial information we can not check abundance gradients in most GRB hosts, and assume the oxygen abundances found are 
representative for the galaxy as a whole, including the progenitor locus. However, Hammer et al.~(2006) and 
Sollerman et al.~(2006) have shown that in the case of 
the host of SN\,1998bw strong differences in $T_e$ and abundances are observed as well.

\subsection{Star formation \label{sfr}}
The detection of the bright \ion{H}{$\alpha$} $\lambda$6563 emission line in the FORS spectrum allows us to accurately 
measure star formation in the host: the \ion{H}{$\alpha$} line luminosity only weakly depends on 
the physical conditions of the ionized gas. 
We use 
\begin{equation}
{\rm SFR}_{\ion{H}{\alpha}} = 7.9 \times 10^{-42} {\rm L}_{\rm H \alpha},
\end{equation}
as found by Kennicutt (\cite{Kennicutt}). Moustakas et al.~(\cite{moustakas}) assess the accuracy of this 
expression through a direct comparison of extinction corrected ${\rm L}_{\rm H \alpha}$ SFR values and
the SFR derived from L$_{\rm IR}$ measurements of a sample of IRAS detected galaxies. They find that when the \ion{H}{$\alpha$}
flux is extinction corrected, the IR and \ion{H}{$\alpha$} SFRs agree without systematic offset with a precision of $\sim$70\%.
The extinction corrected flux of H$\alpha$ can be found from the FORS spectra, yielding
SFR$_{\rm H \alpha}$ = 0.065 $\pm$ 0.005 M$_{\sun}$\,yr$^{-1}$.
Due to slit losses the true \ion{H}{$\alpha$} flux is likely to be higher, and the SFR$_{\rm H \alpha}$ can be interpreted as a lower limit.

Radio and submillimetre observations do not suffer from dust extinction. The radio continuum flux of a normal galaxy (i.e.~non-AGN hosting) 
is thought to be formed by synchrotron emission by accelerated electrons in supernova remnants and by free-free emission from
\ion{H}{II} regions (Condon (\cite{Condon}). It is expected that the radio continuum flux is a particularly good tracer of the recent
SFR, due to the short expected lifetime of the supernova remnants, which is $\lesssim 10^{8}$ yr. We use the method described by
Vreeswijk et al.~(\cite{Vreeswijk01}) and Berger et al.~(\cite{BergerSFR}) to calculate an upper limit to the full star 
formation rate (ie not influenced by any form of dust extinction).
We use the deepest 6 cm (4.9 GHz) Westerbork Synthesis Radio Telescope (WSRT) flux limit, 
i.e.\ when the initial radio afterglow has faded beyond detection limit, and find a 3$\sigma$ limit of 
72 $\mu$Jy (formal flux measurement 8 $\pm$ 24 $\mu$Jy, Kaneko et al.~2006) at a 12h 
full synthesis on April 1 2006. 
This leads to a 3$\sigma$ SFR upper limit of
SFR$_{\rm radio} < 0.15$ M$_{\sun}$\,yr$^{-1}$, which excludes a large amount of obscured star formation when compared to SFR$_{\rm H \alpha}$.
However, we note that despite many similarities, GRB hosts studied to date are a diverse population:
a handful of hosts show clear indications of
much higher star formation rates than seen from optical indicators (comparable to the submillimetre galaxies
at several hundred M$_{\sun}$\,yr$^{-1}$) from
their submm fluxes (Berger et al. 2003; Barnard et al.~2003)
and in one case
(GRB~980703 at $z=0.97$, Berger et al.~2003) its radio flux.  
However, the radio derived SFRs may systematically overestimate star formation through
a non-negligible contribution from AGN activity to the total radio continuum flux.

One of the properties that sets GRB hosts apart from field galaxies is the
specific star formation rate (SSFR): the star-formation rate per unit mass. 
This quantity is shown to be high in GRB hosts (e.g.\ Christensen et al.~2004; Courty et al.~2004). 
Sollerman et al.~(2006) find that the host of GRB\,060218 has $L = 0.008 L_{B}^{*}$, implying
a SSFR of $\sim8$ M$_{\sun}$\,yr$^{-1}$$\left(L/L_{*}\right)^{-1}$. This value is comparable to the
SSFRs for a sample of GRB hosts determined through SED fitting by Christensen et al.~(2004).
Given the multi band SDSS photometry of the host, a low stellar mass is expected.
After modeling the SED of the host, Savaglio, Glazebrook \& Le Borgne (in
preparation) find  $M_\ast=10^{7.2\pm0.3}$ M$_{\sun}$. The measured
metallicity and stellar mass therefore place this galaxy on the
mass-metallicity relation found recently in local dwarf galaxies by Lee
et al.\ (2006).

\section{Discrete velocity components in emission and absorption}
\subsection{Emission line profiles \label{profilesection}}
\begin{figure}
   \centering
   \includegraphics[width=9cm]{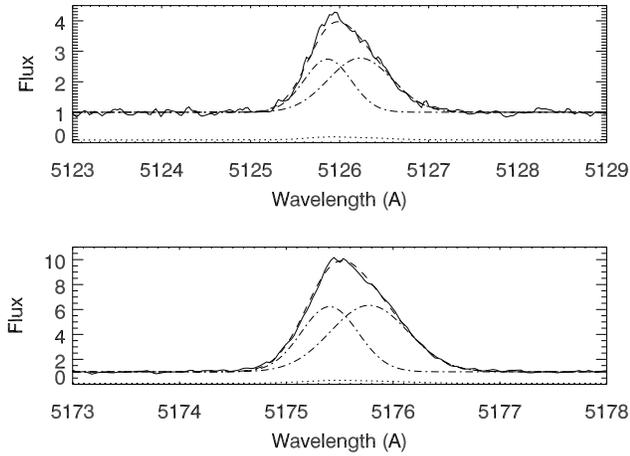}
      \caption{The [\ion{O}{III}] $\lambda$4959 (top) and $\lambda$5008 (bottom) emission lines. The continuum is normalized.
      The error spectrum is shown by a dotted line. The lines are simultaneously fit using fixed flux ratios between $\lambda$4959 and 
      $\lambda$5008 components of 1:3. The two seperate Gaussian components are shown with dash-dotted lines, and the total emission with a dashed line.   
              }
         \label{profileplot}
   \end{figure}

The resolution of the UVES spectrum is sufficient to search for 
structure in the emission line profiles.
Figure \ref{profileplot} shows that the brightest emission lines of [\ion{O}{III}] $\lambda\lambda$4959, 5008 significantly
deviate from a single Gaussian line profile, and are skewed towards the blue. We can rule out an 
instrumental effect as no such effect
was seen in the arc or sky line profiles. Due to a lower signal to noise, we are not able to verify quantitatively whether other emission lines
share the same line profile. We fit Gaussian components to the lines, where the number of components and the
position and width of the components are free parameters. The [\ion{O}{III}] $\lambda\lambda$4959, 5008 lines are fit simultaneously with a fixed flux ratio
between the $\lambda$4959 and $\lambda$5008 components of 1:3. We use both a fit by eye with IRAF {\em splot} and a quantitative deblending using 
VPFIT\footnote{See {\tt http://www.ast.cam.ac.uk/$\sim$rfc/vpfit.html}}. 
Results agree, and we find a good fit (reduced $\chi^2 \sim$  0.72) using two Gaussian components, 
shown in Fig.~\ref{profileplot}, where the components are bluewards and redwards of the average host galaxy redshift. 
For the red component we find a FWHM of 49 $\pm$ 5 km s$^{-1}$ and redshift 0.033453 $\pm$ 0.000019; and
for the blue component a FWHM of 35 $\pm$ 3 km s$^{-1}$ and redshift 0.033379 $\pm$ 0.000005.
This gives ${\Delta}z$ = 7.4 $\pm 2.4 \times 10^{-5}$, or $\sim$21.6 km s$^{-1}$ velocity separation.
These values are similar to those found from spatially resolved high resolution spectroscopy of
emission line regions (i.e.\ the 30 Doradus nebula, see Melnick et al.~1999 for a H$\alpha$ study). 
Arsenault \& Roy (1986) show that $\approx$ 66\% of all giant extragalactic \ion{H}{II} regions show symmetrical H$\alpha$ lines in their integrated
spectra; the remaining profiles show asymmetries similar to the ones in the host of GRB\,060218.
When asymmetric line profiles are observed (i.e.\ in high resolution studies of galactic and extragalactic \ion{H}{II} regions) 
the explanation of their profiles is usually only possible by using a high spatial resolution and 
correlation of the spectra with images. 
As an example, the {\em integrated} H$\alpha$ line profile of 30 Doradus shows a broad and narrow Gaussian component (Melnick et al.~1999), whose
central wavelengths coincide, while spectra taken at multiple positions in the nebula show a large variety of line profiles and 
Gaussian components,
which generally can be associated with filaments in the nebula. 
In the case of this host galaxy, we observe the integrated profile over 7 kpc. 

There are several possible explanations for the occurence of two lines, 
with the broader one redshifted with respect to the 
blue component. One possibility is that we see two separate star forming regions in the host. An alternative explanation is  that the 
components are caused by an expanding shell (bubble) around the starforming region; or by infalling
gas onto the H II region (e.g. ionised gas ejected by perhaps SN shocks or stellar winds, that falls back). Neither of the last two scenarios seem 
plausible: an expanding shell is not likely to have the measured velocity width; and the infalling gas would have to be very highly excited to 
generate the required [\ion{O}{III}] luminosity, which makes it difficult to have co-exisiting \ion{Na}{I} (see \S4.2).
We tentatively interpret the two components as arising from two different starforming regions in the host. High
resolution imaging would be necessary to confirm this. There are several GRB hosts where HST images resolve the host into multiple 
starforming regions with similar
brightnesses (see e.g.~Fruchter et al.~2006).  
We note that this is the first identification of resolved emission line components in a GRB host galaxy spectrum.

\subsection{Absorption lines in the circumburst medium}
\begin{table*}[t]
\caption{Absorption lines in the UVES spectrum of GRB~060218}
\label{t1_sa}
\begin{center}
\begin{tabular}{clcccccc}
\hline
System & Line &  $\lambda_o$ & $z$ & $W_o$ & $\log N$ & $b$ & $\log$ (\ion{Na}{I}/\ion{Ca}{II}) \\
& & (\AA) & & (\AA) & [cm$^{-2}$] & (km s$^{-1}$) & \\
\hline
1 & \ion{Na}{I} $\lambda$5891 & 6088.23 & 0.033378 & 0.091 $\pm$ 0.008 &
$11.79\pm0.04$ & $6.3\pm1.0$ & $-0.54\pm0.08$ \\
 & \ion{Na}{I} $\lambda$5897 & 6094.41 & & 0.049 $\pm$ 0.007 & & & \\
 & \ion{Ca}{II}    $\lambda$3934 & 4066.05 & 0.033363 & 0.093 $\pm$ 0.029 &
$12.33\pm0.07$ & $13.3\pm3.2$ & \\
 & \ion{Ca}{II}    $\lambda$3969 & 4102.03 & & 0.064 $\pm$ 0.028 & && \\
\hline
2 & \ion{Na}{I} $\lambda$5891 & 6088.73 & 0.033462 & 0.071 $\pm$ 0.006 &
$12.22\pm0.22$ & $1.1\pm0.2$ & $0.01\pm0.23$ \\
 & \ion{Na}{I} $\lambda$5897 & 6094.90 & & 0.064 $\pm$ 0.006 & & \\
 & \ion{Ca}{II}    $\lambda$3934 & 4066.42 & 0.033458 & 0.075 $\pm$ 0.025 &
$12.21\pm0.08$ & $4.6\pm1.8$ \\
 & \ion{Ca}{II}    $\lambda$3969 & 4102.40 & & 0.054 $\pm$ 0.029 & & \\
\hline
Galaxy & \ion{Na}{I}  $\lambda$5891 & 5891.75 & 0.0000285 & 0.33 & $12.68\pm0.02$
& $6.4\pm0.2$ & $0.27\pm 0.06$ \\
   & \ion{Na}{I}  $\lambda$5897 & 5897.73 &           & 0.25 & & & \\
   & \ion{Ca}{II}  $\lambda$3934 & 3934.93 & 0.0000381 & 0.13 & $12.41\pm0.06$
& (5) & \\
   & \ion{Ca}{II}  $\lambda$3969 & 3969.74 &           & 0.08 & & & \\
\hline
\end{tabular}
\end{center}
\end{table*}

 \begin{figure}[ht]
   \centering
   \includegraphics[width=7.8cm]{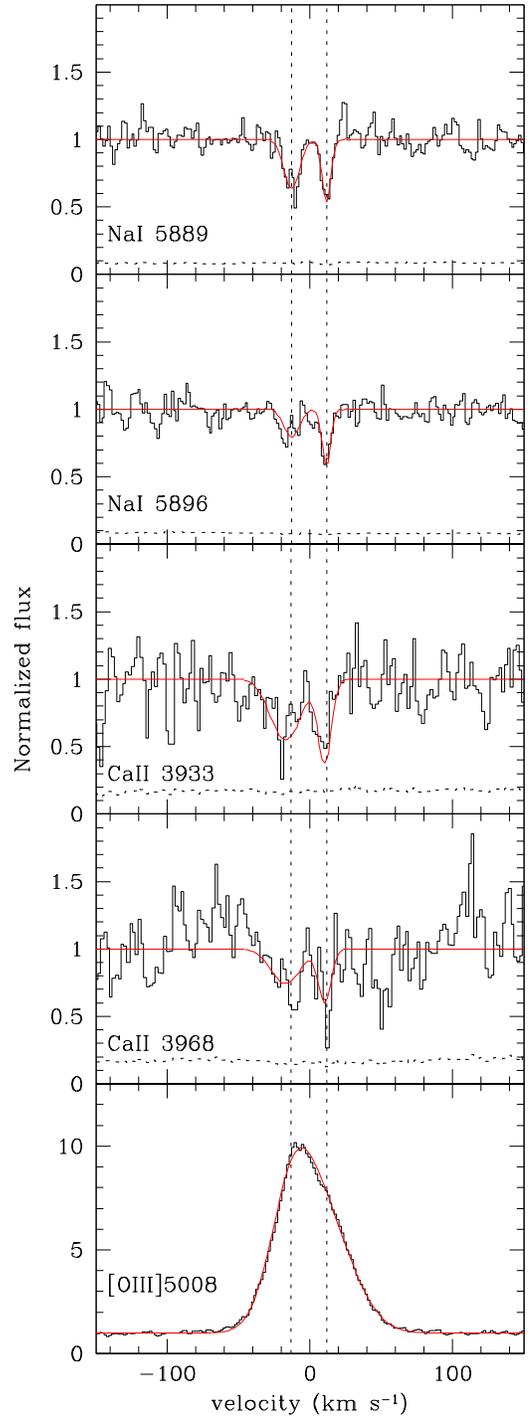}
   \caption{The \ion{Na}{I} and \ion{Ca}{II} absorption lines detected in GRB~060218. The
     two velocity components are marked by the dashed lines (more precisely, these
     mark the position of the two \ion{Na}{I} absorption lines). The smooth line is
     the result of the best fit Voigt profile (reported in
     Table~\ref{t1_sa}). The bottom panel shows the [OIII] $\lambda5008$
     emission line for comparison. The dotted spectrum is the noise.}
    \label{fig1_sa}
  \end{figure}

\begin{figure}
\centering
\includegraphics[width=7.5cm]{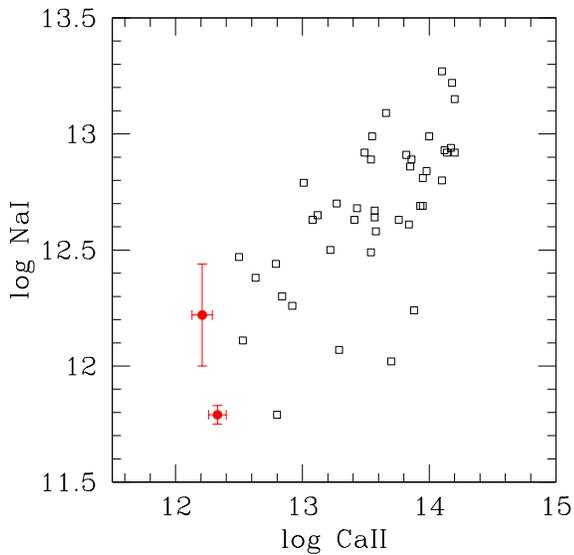}
\caption{\ion{Na}{I} and \ion{Ca}{II} column densities for the two systems in the host of GRB~060218 (filled dots)
and lines of sight in the Milky Way (open squares; typical error for these cases is
$<0.1$ dex; Hunter et al.\ 2006).}
\label{fig4_sa}
\end{figure}

In the UVES spectrum we detect absorption lines at $\sim4070$ \AA,
$\sim4100$ \AA\ and $\sim6090$ \AA, associated with
\ion{Ca}{II}\,$\lambda\lambda3934,3969$ and
\ion{Na}{I}\,$\lambda\lambda5891,5897$ in the foreground gas of the
GRB\,/\,SN, and within the host galaxy.  Figure~\ref{fig1_sa} shows at
least two discrete velocity components, separated in velocity by
$\sim$24 km s$^{-1}$ (systems 1 and 2 in Table~\ref{t1_sa}; see also
Guenther et al.\ 2006). The \ion{Ti}{II} absorption lines are not
detected because they are in a very noisy region of the spectrum
($\lambda<3500$ \AA).

The observed lines have been fitted with Voigt profiles using the
MIDAS package FITLYMAN (Fontana \& Ballester 1995). Figure~\ref{fig1_sa} shows the fitting results, and column
densities and Doppler parameters are listed in Table~\ref{t1_sa}.  The
lines are barely resolved, i.e.\ the line widths are at the level of
the UVES spectral resolution FWHM $\simeq 6.5$ km s$^{-1}$.

The measured FWHM corresponds to an instrumental PSF of 3.9 km s$^{-1}$,
if expressed in terms of Doppler parameter, which can be translated 
into an upper limit on the gas
temperature (derived from the lightest element Na, assuming thermal
broadening) of $T \lesssim 2\times10^4$ K.  Narrow metal lines are
often detected in GRB afterglows when high resolution spectra are
acquired (see for instance Chen, Prochaska, \& Bloom 2006).

There are indications that \ion{Na}{I} and
\ion{Ca}{II} are not tracing each other. The
positions of the two
\ion{Ca}{II} components are shifted by a few km s$^{-1}$ with
respect to \ion{Na}{I}, which could be consistent with uncertainties
on the fitted parameters (more severe for the \ion{Ca}{II}
doublet). However, the line broadenings and \ion{Ca}{II}/\ion{Na}{I}
ratios are different in the two components.  We note that in the
Galactic ISM
\ion{Ca}{II} and \ion{Na}{I} are found in regions with different
physical conditions. The former is found to be generally broader than
the latter, indicating that it traces warmer, more turbulent, and/or
larger gas clouds (Welty et al.\ 1996).

Remarkably, the positions of the two
absorption systems are consistent with the redshifts of the two emission-line
components in the \ion{H}{II} regions derived independently (see
\S3.5 and the lower panel of Figure~\ref{fig1_sa}). The difference in
redshift between emission and absorption is $<1$ km s$^{-1}$ and
$\sim2$ km s$^{-1}$ for the blueshifted and redshifted systems,
respectively. The relative velocity for the two emission components is
21 km s$^{-1}$, close to the 24 km s$^{-1}$ measured from the
absorption systems. However, the broader component in absorption is at
the lowest redshift, whereas the opposite is true for the emission.

The \ion{Ca}{II} column density was measured in another two GRB
afterglows (Savaglio
\& Fall 2004; Savaglio 2006) with a total column density
of nearly $10^{14}$ cm$^{-2}$ in each of them.
\ion{Na}{I} absorption in GRBs is reported here for the first time,
basically due to a lack of suitable data in past GRB observations
(\ion{Na}{I} is redshifted into the NIR for $z>0.7$). The
\ion{Na}{I} and \ion{Ca}{II} abundances have been studied in the Galaxy and LMC
(Hunter et al. 2006; Vladilo et al.\ 1993). Beyond the Local Group,
\ion{Na}{I} and \ion{Ca}{II} have been detected in 2 damped Lyman-$\alpha$
systems (DLAs) in QSO spectra, at
$z=1.062$ and 1.181 (Petitjean et al.~2000; Kondo et al.\ 2006). Other
QSO absorption line studies report upper limits for \ion{Na}{I}
(Boksenberg et al. 1978; 1980). GRB\,060218 is the third source
outside the Local Group where \ion{Na}{I} is measured \footnote{More
\ion{Na}{I} lines are detected in low $z$ SDSS QSO spectra, but no
column density measurements are reported for these systems (V.\ Wild,
private communication.)}.

It is rather complicated to interpret the detection of \ion{Na}{I} and
\ion{Ca}{II} in GRB~060218 in terms of relative abundances. The
ionization potentials of the two ions are quite different (5.1 eV and
11.9 eV for \ion{Na}{I} and \ion{Ca}{II}, respectively). 
\ion{Na}{II} is likely the dominant ion in a neutral-gas environment
(the
\ion{H}{I} ionization potential is 13.6 eV), whilst \ion{Ca}{III}
dominates the calcium species.  To derive the total
abundance of calcium, a significant ionization correction can therefore
be necessary.
Moreover, Ca is an $\alpha$ element, while Na is not. Hence
in low metallicity systems, like GRB\,060218, an $\alpha$-element
enhancement is expected. Sodium was found to trace iron in stars with
metallicities close to Solar, but it can have lower abundances for
lower metallicities (Timmes et al.~1995).  Our
expectations are further complicated by the rather different
refractory properties of the two elements: Na is little
depleted on to dust grains, whereas Ca can be 99\% depleted (Welty et
al.\ 1994; Savage \& Sembach 1996). This problem may be somewhat
 mitigated by the
likely negligible dust depletion in the gas, as suggested by the small
dust extinction derived in the \ion{H}{II} regions of GRB~060218 (see
\S3.1).

Nevertheless, we compare \ion{Na}{I} and \ion{Ca}{II} in GRB~060218
with what is typically observed in the ISM of the Milky Way
(Figure~\ref{fig4_sa}).  The two systems in the host of GRB\,060218
lie in the bottom left corner of the distribution. Is the observed Na
and Ca behaving like the neutral gas of the Milky Way? If the
absorption lines are arising in a neutral region of the ISM, and if we
consider the empirical relation that links \ion{H}{I} to
\ion{Na}{I} (derived by Hunter et al.\ 2006) 
we would expect an \ion{H}{I} column density along the GRB sight line
of the order of $\log N_{\ion{H}{I}}=20.6$ cm$^{-2}$. However, if the
metallicity in the neutral gas is similar to the \ion{H}{II} regions probed
by the emission lines, a much lower total \ion{H}{I} column density is
more likely: $\log N_{\ion{H}{I}}\sim19.4$, suggesting that perhaps
the \ion{Na}{I}-\ion{H}{I} relation found for the Milky Way may not
be applicable for the gas
around GRB\,060218
or that the metallicity in the galaxy is not uniform.
An \ion{H}{I} column density
of the order of 10$^{19.4}$ cm$^{-2}$ is relatively
large, considering that the stellar mass of this GRB host is more than
1000 times smaller
than that of the Milky Way. However, a large reservoir of gas is
expected given the high SFR per unit stellar mass estimated for the host (see section \ref{sfr}).

\section{Secondary metallicity calibrators}
The association of GRB\,060218 with a low mass, low metallicity, high excitation host galaxy follows the trend seen from the other 
nearby GRBs (e.g. Sollerman et al.~2005). In fact, the very low metallicity of the host significantly extends the known 
metallicities of GRB hosts through emission line spectroscopy. 
If the redshift of the galaxy had been substantially higher, the [\ion{O}{III}] $\lambda$4364 line would have gone undetected, making
a direct determination of the abundance through $T_e$ impossible, and 
the metallicity inferred from secondary calibrators would have placed the host at significantly higher metallicity 
(e.g. Mirabal et al.~2006; Modjaz et al.~2006). 
The sample of GRB host galaxies with a measure of abundances through $T_e$ is limited to just four nearby sources 
 due to the faintness of the  [\ion{O}{III}] $\lambda$4364 line, where GRB\,020903 has the highest redshift with $z = 0.25$.
For all other galaxies we are forced to use secondary, empirical methods to calculate metallicity. The most common  
is the R23 calibrator (see e.g. Kobulnicky \& Kewley 2004), which uses the bright [\ion{O}{III}], H$\beta$ and [\ion{O}{II}] lines. 
This method produces a degenerate metallicity solution, which can be broken through other emission lines (e.g. [\ion{N}{II}] and H$\alpha$), 
but in many cases these other lines are not available (due to e.g. redshift or insufficient S/N). The R23 method is calibrated through 
photo-ionization models, which have limitations at the low and high metallicity ends. It has been shown that R23 metallicities 
have an offset with respect to $T_e$ metallicities for metallicities close to Solar (e.g. Bresolin et al.~2004). 
Modjaz et al.~(2006) use the R23 method to find 12 + log(O/H) = 8.0 $\pm$ 0.1 for the host of GRB\,060218.
An alternative method that is calibrated on a sample of H\,II regions with $T_e$ determined metallicities,
is the ratio of the nebular [\ion{O}{III}] and [\ion{N}{II}] lines (e.g. Pettini \& Pagel 2004). 
The $N2$ index ($N2 = {\rm log}([\ion{N}{II}]\lambda6583/{\rm H}\alpha)$) has been proposed particularly for low metallicity
galaxies. However, for GRB 060218, the $N2$ calibration
overestimates the metallicity of the host by more than a factor of 3, 
yielding 12 + log(O/H) $\sim$ 8.2, but the scatter in this relation is large 
(90\% of the fitted data in Pettini \& Pagel 2004 falls within a range log(O/H) = $\pm$0.4). 
The $O3N2 = {\rm log}\left(([\ion{O}{III}]\lambda5008/{\rm H}\beta)/([\ion{N}{II}]\lambda6583/{\rm H}\alpha) \right)$ ratio has significantly less
scatter but above $O3N2 \gtrsim 1.9$ this indicator breaks down and cannot be reliably used (Pettini \& Pagel 2004). Our host 
has $O3N2$ = 1.88 $\pm$ 0.11, which may explain
the overestimate of the metallicity through this indicator of 12 + log(O/H) $\approx$ 8.1. 

Concluding, we can state that an analysis using a variety of different secondary methods (e.g. R23, $O3N2$) would not have found the true metallicity, 
but may still have identified this host as a very low metallicity candidate.

\section{Massive stars and progenitors}
Long GRB progenitors are likely massive Wolf-Rayet (WR) stars. We can gain a greater understanding of the evolutionary paths of such massive 
stars towards GRBs by detecting WR populations within GRB host galaxies.
A valuable diagnostic on the 
massive star population is the \ion{He}{II} $\lambda$4686 line, which appears as a broad line and/or as a nebular line
in some \ion{H}{II} galaxies (Schaerer et al.~1999), and is a direct sign of (unusually) high excitation levels caused by the presence of WR stars 
(especially WC- and WO-type WR stars).
Together with, amongst others, \ion{N}{III} $\lambda$4640, \ion{C}{IV} $\lambda$4658, [\ion{Fe}{III}] $\lambda$4658 and 
[\ion{Ar}{IV}] $\lambda$4711 lines, this line forms the so-called blue WR bump (from $\sim$4650 -- 4700 \AA) in low resolution spectra, 
which is often accompanied by a \ion{C}{IV} $\lambda$5808 line 
(the red bump). 
In a recent deep spectroscopic search in nearby GRB host galaxies, Hammer et al.~(2006) have detected the \ion{He}{II} line and accompanying
WR bump in the spectra of the hosts of GRB\,980425, 020903 and 031203. 
From the relative intensities of the H $\beta$ 
and \ion{He}{II} $\lambda$4686 lines or WR bump (measured flux ratios H $\beta$\,/\,\ion{He}{II} $\lambda$4686 generally range from $\sim$0.01-0.02) one 
can estimate the ratio of WR to O stars.

In Crowther \& Hadfield (2005) the effect of metallicity on the derived WR\,/\,(WR + O) 
ratio is investigated, by comparing the fluxes of SMC, LMC and Galactic WR stars with atmosphere models. 
They find a lower WR line luminosity at decreasing metallicity. 
In the hosts of GRBs\,980425 and 020903, Hammer et al.~(2006) find values of
WR\,/\,O $\sim0.05$ and $\sim0.14 - 0.2$, respectively. We note that no metallicity correction has been applied 
to these values, which would increase the number of WR stars as the studied hosts 
have sub-Galactic metallicity.

These high WR\,/\,O ratios are of particular interest as their 
metallicities are accurately known from $T_e$ analyses: 0.5 $Z_{\sun}$ 
and 0.19 $Z_{\sun}$ for GRB\,980425 and GRB\,020903, respectively (Hammer et al.~2006).
At these low metallicities the high measured WR\,/\,O ratios may be regarded as evidence for very recent star bursts,
 but it is difficult to determine the ages of the dominant stellar populations.
Optical and near-infrared SED fitting has shown that the dominant stellar populations 
in GRB hosts are young at typically $\sim$0.1 Gyr (Christensen et al.~2004). 
In the case of the host of GRB\,980425 the entire galaxy is well fitted with a continuous star formation history (Sollerman et al.~2005),
although Hammer et al.~(2006) find a young population from the equivalent widths of the emission lines.
No strong evidence for a very young starburst is apparent in the host spectrum of GRB\,020903, 
but the merger morphology of the host and the emission line EWs may suggest that some recent star formation is present. 
However, the WR/O ratio of 0.14 -- 0.2 in the host of GRB\,020903 is particularly remarkable, 
as such abundant production of WR stars at the low metallicity of 0.19 $Z_{\sun}$ can only
be explained by invoking instant star bursts with peculiar initial mass function 
within the standard star burst model by Schaerer \& Vacca (1998) (e.g.~Fernandes et al.~2004). 

Recent stellar evolution models indicate that the effects of rotation may be, in part, 
responsible for observed high WR\,/\,O ratios in galaxies.  According to Meynet \& Maeder (2005a), 
including the effects of rotation significantly enhances the mass loss rates of massive stars 
during the giant phase compared to the non-rotating case, as the shear instability due to 
the strong degree of differential rotation between the core and the envelope induces fast chemical mixing. 
Their models predict a WR\,/\,O ratio of about 0.02 at $Z = 0.004$ for a constant star formation, and in principle 
WR\,/\,O $\sim$ 0.2 might be achieved at the given metallicity for instant star bursts even with a standard
initial mass function.
However, such rotating models predict GRB\,/\,SN\,Ibc ratios that are too high.  
Their prediction of spin rates of young neutron stars is also inconsistent with observations
(Hirschi, Meynet \& Maeder 2005; see also Heger, Langer \& Woosley 2000).

More recent models that include magnetic torques for the angular momentum transport in the star
suggest another way to produce WR stars at low metallicity. 
Although in magnetic models the chemical mixing induced by the shear instability during the giant phase is
negligible as the magnetic torque tends
to keep the star rigidly rotating, the mixing by meridional circulation can be
very fast even on the main sequence (Maeder \& Meynet 2005b). This may even cause
the whole star to become chemically homogeneous if the initial rotational
velocity is sufficiently high, especially at sub-solar metallicity
(Yoon \& Langer 2005; Woosley \& Heger 2006).

Formation of WR stars can thus be induced not only by mass loss but also by chemical mixing, and
Woosley \& Heger (2006) found that some WR stars formed through such 
chemically homogeneous evolution 
can actually retain enough angular momentum necessary for GRB production.
The predicted GRB\,/\,SN ratio through such evolutionary channels also turns out to 
be consistent with observations, when the observationally derived initial rotational velocity distribution of 
massive stars by Mokiem et al.~(2006) is adopted (Yoon, Langer \& Norman 2006).
Interestingly, as stars may be kept rotating rapidly at low metallicity due to lowered mass loss rates 
(e.g.~Vink et al.~2005), this new type of evolution by fast rotational mixing can lead to high WR\,/\,O ratios 
even at very low metallicity (Yoon, Langer \& Norman 2006).
It also predicts rather large delay times for WR production from the star formation 
(i.e., from several $10^6$ yrs to a few $\sim10^7$ yrs; Yoon, Langer \& Norman 2006) compared to 
those from mass-loss induced WR formation ( $<$ a few $10^6$ yrs).

In this regard, an estimate of the WR\,/\,O ratio in the host galaxy of GRB\,060218 
might provide a valuable test case for the new GRB progenitor scenario by Yoon \& Langer (2005)
and Woosley \& Heger (2006). 
That is, if the WR\,/\,O ratio in this host turns out to be high despite its very low metallicity, 
it may support the chemically homogeneous evolution scenario for GRB progenitors and extend the 
metallicity -- WR\,/\,O star ratio relation for GRB hosts down to lower metallicity. This relation is well known
for the Milky Way, LMC and SMC, and a deviation of GRB hosts away from that trend gives strong input for progenitor modelling.

The high resolution UVES spectrum of GRB\,060218 should be able to 
resolve the components comprising the WR bump, and resolve a \ion{He}{II} line into a broad and a nebular component. However, no nebular \ion{He}{II} line
or WR bump has been significantly detected.
We measure the flux upper limit on the WR bump in the UVES spectrum by summing the flux in the WR bump wavelength region (restframe 4650 - 4686\AA). 
Following the method of Schaerer \& Vacca (1998) we set an upper limit of WR\,/\,(WR + O) $\lesssim$ 0.4, which is not a constraining limit, 
owing to the fact that the SN outshines the possible \ion{He}{II} line. 
To reliably detect the WR bump in this host we need to detect the host continuum with reasonable S/N, which means the
supernova has to fade below this level before more sensitive searches are feasible.

\section{Conclusions}
We present a VLT UVES high resolution spectrum of SN\,2006aj, associated with the nearby GRB\,060218 at heliocentric redshift $z = 0.03342(2)$. 
We use the emission lines of the UVES spectrum as well as the line measurements from
our FORS spectroscopic campaign to derive properties of the host. We find that the electron density is low, 
and the electron temperature is high, $T_e ({\rm O}^{2+}) = 2.48 ^{+0.5}_{-0.3}  \times 10^4$K, as shown in Table 1. We find a low host
metallicity of 12 + log(O/H) = $7.54^{+0.17}_{-0.10}$, placing it among the most metal
deficient subset of emission-line galaxies. It is also the lowest
metallicity found so far for a GRB host from an emission line analysis.
The metallicity we find lies considerably below the values derived using secondary calibrators, 
e.g.~the metallicity 12 + log(O/H) = 8.0 as derived from the R23 calibrator (e.g.~Modjaz et al.~2006).
The mass of the galaxy is low, 
and matches what is expected from the mass-metallicity relation for dwarf galaxies. 
We measure a relatively high value for log(N/O) with respect to the metallicity, which is also seen in a few other GRB hosts. As our
uncertainty on log(N/O) is relatively high, deeper spectroscopy is needed to confirm this overabundance.

The bright emission lines show strong evidence for asymmetry, and a single Gaussian provides a poor fit to the profiles of the bright
[\ion{O}{III}] emission lines. A two Gaussian model
provides a satisfactory fit with the two components separated by $\sim$22 km s$^{-1}$. We find the same two velocity components in absorption
through the \ion{Ca}{II} and \ion{Na}{I} absorption lines in the host. We tentatively interpret these two velocity components to be due to two
star forming regions in the host galaxy. However, to unravel their true identity, high spatial resolution imaging is needed.

The dust content of the galaxy is low, based on the Balmer line decrement. This is also evident from the low limit on the obscured star 
formation rate we set through a 3$\sigma$ upper limit on the flux at 6 cm of SFR$_{\rm radio} < 0.15$ M$_{\sun}$\,yr$^{-1}$, compared to the
optical star-formation rate SFR$_{\rm H \alpha}$ = 0.065 $\pm$ 0.005 M$_{\sun}$\,yr$^{-1}$. 

This host galaxy is an interesting target for future spectroscopy targeted at the
WR bump, as the low metallicity of the host will significantly extend the present sample of GRB hosts with known WR star content and metallicity.
We show that a measure of these two quantities for a sample of GRB hosts may provide further insight into the nature of GRB progenitors.

The absolute magnitude of the host ($M_B = -15.9$, e.g.~Sollerman et al.~2006) is such that this galaxy would not have been detected 
in any survey at a redshift of $z \sim1$, let alone at the mean {\em Swift} GRB redshift of $z\sim2.8$ (Jakobsson et al.~2006), which makes 
this host an important object to study in the context of larger redshift GRB hosts.

\begin{table*}[t] 
\caption{Table of emission lines used in this paper. Fluxes and EWs are as observed. Fluxes from the FORS spectrum are taken 
from Pian et al.~(2006).}
\label{linestable} 
\begin{center} 
\begin{tabular}{lcc}
\hline
\multicolumn{3}{c}{UVES spectrum} \\
\hline 
ID & EW & Flux \\ 
 & ({\AA})   &  ($\times 10^{-17}$ erg s$^{-1}$ cm$^{-2}$)  \\ 
\hline 
$[\ion{O}{II}]$ $\lambda$3727 & 1.59 $\pm$  0.09 & 75.23 $\pm$ 3.42  \\
$[\ion{O}{II}]$ $\lambda$3729 & 2.62 $\pm$  0.12  & 121.1 $\pm$ 4.0   \\
$[\ion{Ne}{III}]$ $\lambda$3869 & 0.75 $\pm$ 0.08  & 34.06 $\pm$ 3.51   \\
$[\ion{Ne}{III}]$ $\lambda$3969 & 0.26$\pm$ 0.06  & 11.80 $\pm$ 2.64   \\
H$\beta$ $\lambda$4862 & 2.17 $\pm$ 0.06  & 92.92 $\pm$ 1.93	 \\
H$\gamma$ $\lambda$4341 & 0.90 $\pm$ 0.07  &  47.99 $\pm$ 3.46   \\
H$\delta$ $\lambda$4102 & 0.61 $\pm$ 0.07  &  27.83 $\pm$ 3.04    \\
H$\epsilon$ $\lambda$3971 & 0.29 $\pm$ 0.07  & 13.03 $\pm$ 3.02    \\
H8 $\lambda$3889 & 0.27 $\pm$ 0.07  & 12.25 $\pm$  2.90    \\
H9 $\lambda$3835 & 0.18 $\pm$ 0.07  & 7.65 $\pm$ 2.92    \\
$[\ion{O}{III}]$ $\lambda$4960 & 2.88 $\pm$ 0.06  &  139.0 $\pm$ 2.08  \\
$[\ion{O}{III}]$ $\lambda$5008 & 8.34 $\pm$ 0.09  &  426.2 $\pm$ 2.48    \\
$\ion{He}{I}$ $\lambda$5877 & 0.24 $\pm$ 0.04 & 9.55 $\pm$  1.17    \\
\hline
\multicolumn{3}{c}{FORS spectrum} \\
\hline 
 ID &  & Flux \\ 
 &    &  ($\times 10^{-17}$ erg s$^{-1}$ cm$^{-2}$)  \\ 
\hline 
$[\ion{O}{II}]$ $\lambda$3727, 3729  & & 190 $\pm$ 50 \\
(Doublet unresolved)&&\\
$[\ion{Ne}{III}]$ $\lambda$3869&& 31 $\pm$ 5\\
$[\ion{Ne}{III}]$ $\lambda$3969&& 18 $\pm$ 3\\
H$\gamma$ $\lambda$4341 && 37 $\pm$ 6  \\
$[\ion{O}{III}]$ $\lambda$4364 && 21 $\pm$ 4\\
H$\beta$ $\lambda$4862&&101 $\pm$ 4\\
$[\ion{O}{III}]$ $\lambda$4960 &&150 $\pm$ 7\\
$[\ion{O}{III}]$ $\lambda$5008 &&460 $\pm$ 10\\
$\ion{He}{I}$ $\lambda$5877 && 5.5 $\pm$ 1.0\\
H$\alpha$ $\lambda$6563 &&315 $\pm$ 25\\
$[\ion{N}{II}]$ $\lambda$6584 &&19 $\pm$ 2\\
$[\ion{S}{II}]$ $\lambda$6717 &&14 $\pm$ 2\\
$[\ion{S}{II}]$ $\lambda$6731 &&14 $\pm$ 2\\
\hline
\hline 
\end{tabular} 
\end{center} 
\end{table*}

\begin{acknowledgements}
      We thank the observers and Paranal staff for performing the reported observations at ESO VLT. We are very grateful to R.B.C.~Henry, H.~Lee and N.~Tanvir for 
      helpful discussions. We thank the anonymous referee for helpful comments.
      KW thanks NWO for support under grant 639.043.302. The Dark Cosmology Centre is funded by the Danish
National Research Foundation. SCY is supported by the VENI grant (639.041.406) of the Netherlands
Organization for Scientific Research (NWO).
      The authors acknowledge benefits from collaboration within the EU FP5 Research
Training Network ``Gamma-Ray Bursts: An Enigma and a Tool" (HPRN-CT-2002-00294).
\end{acknowledgements}


\begin{thebibliography}{}

\bibitem[2001]{Allendeprieto} {Allende Prieto}, C., {Lambert}, D.~L. \& {Asplund}, M. 2001, 
        ApJ, 556L, 63

\bibitem[1986]{Arsenault} Arsenault, R., \& Roy, J.-R. 1986,
        AJ, 92, 567

\bibitem[2003]{Barnard} Barnard, V.~E., Blain, A.~W., Tanvir, N.~R., et al. 2003,
       MNRAS, 338, 1

\bibitem[2003]{BergerSFR} Berger, E., Cowie, L.~L., Kulkarni, S.~R., et al. 2003,
       ApJ, 588, 99

\bibitem[2002]{bloom} Bloom, J. S., Kulkarni, S. K., \& Djorgovski, S. G. 2002, 
       AJ, 123, 1111

\bibitem[]{} Boksenberg, A., Carswell, R.~F., Smith, M.~G., et al. 1978,
       MNRAS, 184, 773

\bibitem[]{} Boksenberg, A., Danziger, I.~J., Fosbury, R.~A.~E., et al. 1980,
       ApJ, 242, 145  


\bibitem[]{2004} Bresolin, F., Garnett, D.~R. \& Kennicutt, R.~C., Jr. 2004,
       ApJ, 615, 228
	
\bibitem[2006]{campana} Campana, S., Mangano, V.,  Blustin, A.~J., et al. 2006, 
       astro-ph/0603279

\bibitem[1989]{Cardelli} Cardelli, J.~A., Clayton, G.~C. \& Mathis, J.~S. 1989,
         ApJ, 345, 245

\bibitem[2002]{chary} Chary, R., Becklin, E.~E. \& Armus, L. 2002,
       ApJ, 566, 229	
	
\bibitem[]{} 
Chen, H.-W., Prochaska, J.~X., \& Bloom, J.~S.\ 2006, proceedings of
the 16th Annual October Astrophysics Conference in Maryland, "Gamma
Ray Bursts in the Swift Era", eds. S. Holt, N. Gehrels and J. Nousek,
astro-ph/0602144


\bibitem[2005]{chen050730} Chen, H.-W., Prochaska, J.~X., Bloom, et al. 2005
       ApJ, 634, 25

\bibitem[2004]{christensen} Christensen, L., Hjorth, J. \& Gorosabel, J. 2004, 
      A\&A, 425, 913

\bibitem[1992]{Condon} Condon, J.~J. 1992,
       ARA\&A, 30, 575

\bibitem[2004]{Courty} Courty, S., Bj\"ornsson, G. \& Gudmundsson, E.~H. 2004,
       MNRAS, 354, 581

\bibitem[2005]{Crowther} Crowther, P.~A. \& Hadfield, L.~J.  2005,
       A\&A, 449, 711

\bibitem[2000]{Dekker} Dekker, H., D'Odorico, S., Kaufer, A., et al. 2000,
       SPIE, 4008, 534

\bibitem[2006]{EllisonKewley} Ellison, S.~L. \& Kewley, L.~J. 2006,
       Proceedings ``The Fabulous Destiny of Galaxies; Bridging the Past and Present'', 
       astro-ph/0508627

\bibitem[2004]{Fernandes} Fernandes, I.F., de Carvalho, R., Contini, T., et al. 2004, MNRAS, 355, 728

\bibitem[1995]{Fontana} Fontana A. \& Ballester P., 1995, The ESO Messenger, 80, 37

\bibitem[1999]{Fruchter} Fruchter, A.~S., Thorsett, S.~E., Metzger, M.~R., et al. 1999,
       ApJ, 519, L13

\bibitem[2006]{fruchter} Fruchter, A.~S., Levan, A.~J., Strolger, L., et al. 2006, 
       Nat, 441, 463

\bibitem[1995]{fukugita} Fukugita, M., Shimasaku, K. \& Ichikawa, T. 1995,
       PASP, 107, 945

\bibitem[2006]{GCNUVES} Guenther, E.~W., Klose, S., Vreeswijk, P.~M., et al. 2006, 
        GCN Circulars 4863

\bibitem[2006]{HammerWR} Hammer, F., Flores, H., Schaerer, D., et al. 2006,
         A\&A, 454, 103

\bibitem[2000]{heger} Heger, A., Langer, N., \& Woosley, S.E. 2000, ApJ, 528, 368

\bibitem[2005]{Hirschi} Hirschi, R., Meynet, G., \& Maeder, A. 2005, A\&A, 443, 581

\bibitem[2003]{Hjorth} Hjorth, J., Sollerman, J., M\o ller, P. et al. 2003,
          Nat, 423, 847

\bibitem[2006]{Hunter} Hunter, I., Smoker, J.~V., Keenan, F.~P., et al. 2006, 
        MNRAS, 367, 1478

\bibitem[1994]{Izotov} Izotov, Y.~I., Thuan, T.~X. \& Lipovetsky, V.~A. 1994,
       ApJ, 435, 647

\bibitem[1999]{IzotovIZw18} Izotov, Y.~I., Chaffee, F.~H., Foltz, C.~B., et al. 1999,
        ApJ, 527, 757

\bibitem[2006a]{Izotov1} Izotov, Y.~I., Stasi\'{n}ska, G., Meynet, G., et al. 2006a,
        A\&A, 448, 955 

\bibitem[2006b]{Izotov2} Izotov, Y.~I., Papaderos, P., Guseva, N.~G., et al. 2006b,
        A\&A, 454, 137

\bibitem[2006b]{Jakobsson} Jakobsson, P., Bj{\"o}rnsson, G., Fynbo, J.~P.~U., et al. 2005,
        MNRAS, 362, 245

\bibitem[2006]{kaneko} Kaneko, Y., Ramirez-Ruiz, E., Granot, J.,  et al. 2006,
        astro-ph/0607110
	
\bibitem[2003]{kauffmann} Kauffmann, G., Heckman, T.~M., Tremonti, C., et al. 2003,
        MNRAS, 346, 1055

\bibitem[Kawai et al.~2005]{Kawai} Kawai, N., Kosugi G., Aoki, K., et al. 2006,
        Nat, 440, 184

\bibitem[1998]{Kennicutt} Kennicutt, R.~C. 1998,
       ARA\&A, 36, 189 

\bibitem[2006]{} Kewley, L.~J., Brown, W.~R., Geller, M.~J. et al. 2006,
       astro-ph/0609246	
	
\bibitem[2004]{Kobulnicky} Kobulnicky, H.~A. \& Kewley, L.~J. 2004,
       ApJ, 617, 240 
 
\bibitem[2006]{KondoNaCalines} Kondo, S., Kobayashi, N., Minowa, Y., et al. 2006,
       ApJ, 643, 667    

\bibitem[2006]{LangerNorman} Langer, N., \& Norman, C.~A. 2006 
       ApJ, 638, L63

\bibitem[2003]{Lodders} Lodders, K. 2003,
       ApJ, 591, 1220

\bibitem[2006]{} Lee, H., Skillman, E.~D., Cannon, J.~M., et al. 2006, 
       ApJ, in press, astro-ph/0605036

\bibitem[2006]{} Lee, H., Skillman, E.~D. \& Venn, K.~A. 2006,
       ApJ, 642, 813

\bibitem[2003]{lefloch} Le Floc'h, E., Duc, P.-A., Mirabel, I.~F., et al. 2003,
       A\&A, 400, 499

\bibitem[2003]{lopez} Lopez, S., \& Ellison, S.~L.  2003,
       A\&A, 403, 573

\bibitem[2006]{MazzaliNature} Mazzali, P.~A., Deng, J., Nomoto, K., et al. 2006,
        Nat, 442, 1018

\bibitem[1999]{Melnick}	Melnick, J., Tenorio-Tagle, G. \& Terlevich, R. 1999,
        MNRAS, 302, 677

\bibitem[2005]{meynet1} Meynet, G., \& Maeder, A. 2005a, A\&A, 429, 581

\bibitem[2005]{meynet2} Meynet, G., \& Maeder, A. 2005b, A\&A, 440, 104	

\bibitem[2006]{GCNMirabal} Mirabal, N. \& Halpern, J.~P. 2006,
        GCN Circulars 4792

\bibitem[2006]{Mirabalpaper} Mirabal, N., Halpern, J.~P., An, D., et al. 2006,
         ApJ, 643, 99

\bibitem[2006]{Modjaz060218}  Modjaz, M., Stanek, K.~Z., Garnavich, P.~M. et al. 2006,
        ApJ, 645, 21

\bibitem[2006]{rohied} Mokiem, M., de Koter, A., Evans, C. et al., 2006, 
        A\&A, 456, 1131

\bibitem[2006]{moustakas}  Moustakas, J., Kennicutt, R.~C., \& Tremonti, C.~A. 2006,
        ApJ, 642, 775

\bibitem[1989]{Osterbrock} Osterbrock, D.~E. 1989,
       Astrophysics of Gaseous Nebulae and Active Galactic Nuclei (Mill Valley: University Science Books) 

\bibitem[1992]{Pagel} Pagel, B.~E.~J., Simonson, E.~A., Terlevich, R.~J., et al. 1992,
       MNRAS, 255, 325

\bibitem[]{} Petitjean, P., Aracil, B., Srianand, R., et al. 2000, 
       A\&A, 359, 457

\bibitem[2004]{Pettini} Pettini, M. \& Pagel, B.~E.~J. 2004,
       MNRAS, 348, L59
       
\bibitem[2006]{PianNature} Pian, E., Mazzali, P.~A., Masetti, N., et al. 2006,
       Nat, 442, 1011
       		
\bibitem[2004]{Prochaska031203} Prochaska, J.~X., Bloom, J.~S., Chen, H., et al. 2004,
       ApJ, 611, 200

\bibitem[]{} Savage, B. D., \& Sembach, K. R. 1996, 
       ARA\&A, 34, 279

\bibitem[]{} Savaglio, S., \& Fall, S.~M.\ 2004, 
       ApJ, 614, 293 

\bibitem[]{} Savaglio, S., 2006, New Journal of Physics, 8, 195

\bibitem[1998]{schaerer} Schaerer, D. \& Vacca, W.~D. 1998,
      ApJ, 497, 618

\bibitem[]{}Schaerer, D., Contini, T. \& Pindao, M. 1999,
      A\&AS, 136, 35

\bibitem[Schlegel et al.~1998]{Schlegel98} Schlegel, D.~J., Finkbeiner, D.~P. \& Davis, M. 2004,
       ApJ, 500, 525

\bibitem[1994]{Skillman} Skillman, E.~D., Terlevich, R.~J., Kennicutt, R.~C., Jr., et al. 1994,
       ApJ, 431, 172

\bibitem[2006]{Soderberg060218radio} Soderberg, A.~M., Kulkarni, S.~R., Nakar, E., et al. 2006,
       Nat, 442, 1014

\bibitem[2005]{Sollerman3gal}  Sollerman, J., {\"O}stlin, G., Fynbo, J.~P.~U., et al. 2005,
       NewA, 11, 103

\bibitem[2006]{Sollerman060218}  Sollerman, J., Jaunsen, A.~O., Fynbo, J.~P.~U., et al. 2006,
        A\&A, 454, 503
  
\bibitem[2003]{Stanek030329} Stanek, K.~Z., Matheson, T., Garnavich, P.~M., et al. 2003,
        ApJ, 591, 17     

\bibitem[2006]{StanekZ} Stanek, K.~Z., Gnedin, O.~Y., Beacom, J.~F., et al. 2006,
       astro-ph/0604113

\bibitem[2005]{Starling} Starling, R.~L.~C., Vreeswijk, P.~M., Ellison, S.~L., et al. 2005,
       A\&A, 442, 21

\bibitem[2003]{Stasinska} Stasi\'{n}ska \& Izotov  2003,
       A\&A, 397, 71



\bibitem[Timmes et al.(1995)]{1995ApJS...98..617T} Timmes, F.~X., Woosley,
S.~E., \& Weaver, T.~A.\ 1995, \apjs, 98, 617


\bibitem[2001]{lacosmic} Van Dokkum, P.~G. 2001,
       PASP, 113, 1420

\bibitem[2001]{vink1} Vink, J.S.,  de Koter, A., \& Lamers, H.J.G.L.M. 2001, 
        A\&A, 369, 574

\bibitem[1993]{vladilo} Vladilo, G., Centurion, M \& Cassola, C. 1993, 
       A\&A, 273, 3

\bibitem[2001]{Vreeswijk01} Vreeswijk, P.~M., Fender, R.~P., Garrett, M.~A., et al. 2001,
        A\&A, 380, L21


\bibitem[1994]{welty} Welty, D.~E., Hobbs, L.~M., Kulkarni, V.~P.\ 1994, 
         ApJ, 436, 152
	 
\bibitem[1996]{welty2} Welty, D.~E., Morton, D.~C. \& Hobbs, L.M. 1996,
         ApJS, 106, 533	 

\bibitem[2006]{wolf} Wolf, C. \& Podsiadlowski, P. 2006,	
	astro-ph/0606725

\bibitem[2006]{woosley} Woosley, S.~E. \& Heger, A. 2006, 
         ApJ, 637, 914

\bibitem[2005]{Yoon} Yoon, S.-C. \& Langer, N. 2005, 
        A\&A, 443, 643

\bibitem[2006]{Yoon2} Yoon, S.-C., Langer, N., \& Norman, C. 2006, 
       astro-ph/0606637


\end{thebibliography}
\end{document}